\documentclass[11pt]{article}
\usepackage{geometry}
\usepackage{graphicx}
\usepackage{amssymb}

\title{The Hanbury Brown Twiss effect for atomic matter waves}

\author{Christoph I. Westbrook and Denis Boiron\\
{\it Laboratoire Charles Fabry de l'Institut d'Optique, Univ. Paris
Sud, CNRS,}\\
\it{Campus Polytechnique RD128, 91127 Palaiseau France} \\ \\
{Quantum of Quasars workshop - QQ09},\\
December 2-4, 2009,
Grenoble, France}
\date{}

\begin{document}
\maketitle

\begin{abstract}
This paper discusses our recent work on developing the matter wave analogs
to the Hanbury Brown Twiss experiment. 
We discuss experiments using cold atoms, both bosons and fermions, both 
coherent and incoherent. Simple concepts from classical and quantum optics suffice to understand most of the results, but the ideas can also be traced back to the work of Einstein on the thermodynamics of Bose gases. 
\end{abstract}

\section{Introduction}
\begin{figure}[t] 
   \centering
   \includegraphics[width=9.5cm]{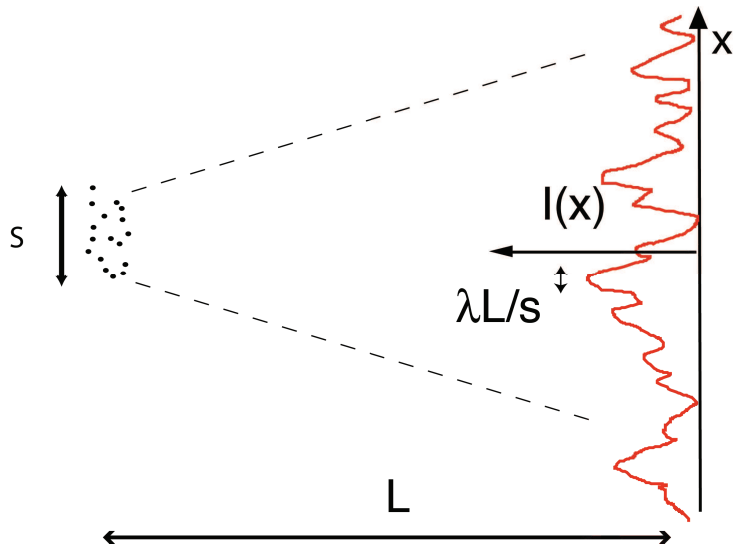} 
   \caption{Speckle interpretation of the HBT effect. Waves from many small sources, spread over a size $s$, interfere at the detector plane at distance $L$. The resulting speckle pattern has a correlation length of $l=\lambda L/2\pi s$. If the phases of the sources fluctuate, the speckle pattern does as well and a detector would record a temporally fluctuating light intensity. The fluctuations at two such detectors are correlated if the detector separation is less than $l$, and uncorrelated otherwise. Thus a correlation measurement permits the determination of $l$ and of $s/L$, the angular size of the source.}
   \label{fig:speckle}
\end{figure}

Experimental physicists usually fight to eliminate noise. 
But the history of physics contains celebrated examples in which the study of noise, or more precisely fluctuation phenomena, has led to significant discoveries. An example is Einstein's analysis of Brownian motion \cite{einstein:05} which played a crucial role in the development of the atomic theory of matter. And several times in his later career, Einstein returned to the study of fluctuations. In the famous 1925 article in which he described Bose-Einstein condensation \cite{einstein:25}, Einstein considered the fluctuation in the number of particles $N$ in a small volume within a larger volume of an ideal gas at a given temperature. Using thermodynamic arguments, he found a formula for the variance of $N$:
\begin{equation}
\delta N^2=\langle N \rangle+\langle N \rangle^2/g
\label{eq:variance}
\end{equation}
Recall that the variance is the mean squared deviation from the mean of $N$: 
$\delta N^2=\langle (N-\langle N \rangle)^2 \rangle$. In Eq.~\ref{eq:variance} the quantity $g$ is the number of phase space cells occupied by the gas, that is the phase space volume divided by Planck's constant cubed: $g = (\Delta x \Delta p/h)^3$.
	In reading the paper, one can sense his fascination with this formula: he identifies the linear term with the fluctuations of independent particles, what we often call "shot noise" today. Einstein recognizes the quadratic term as being due to an interference effect. Indeed he had already made a similar analysis in the case of radiation \cite{einstein:09}, in which case the quadratically increasing variance can be interpereted as "speckle" (see figure~\ref{fig:speckle}). Here however, the formula applies to matter and he remarks on the "mutual influence between the particles of an altogether puzzling nature". With the formulation of the quantum theory in the ensuing few years, it became clear that he had put his finger once again on the wave-particle duality.
	
	This formula became standard fare in textbooks on quantum statistical mechanics, but like many calculations in textbooks, it applied to a {\it Gedankenexperiment}. The quadratic term is usually extremely small compared to the linear term, which itself is often difficult to observe. For example at atmospheric temperature and pressure, 1 mm$^3$ of air contains about $2\times 10^{16}$ molecules. According to Eq.~\ref{eq:variance}, the shot noise is of order $10^{-8}$ and the interference term is another $10^6$ times smaller. In other words, the spatial and temporal coherence lengths of a typical gas are extremely small. To our knowledge the first observation of these interference fluctuations was made using light in the famous experiments of Robert Hanbury Brown and Richard Twiss (HBT) \cite{brown}. Although the effect is easily explained in the context of classical wave optics, as in figure~\ref{fig:speckle}, the HBT experiment stimulated deep questions about the quantum description of light. It is not at all obvious how two photons, emitted from  widely separated points on a source as incoherent as a star could conspire to arrive at two detectors on earth in a correlated way. 
An argument due to Fano \cite{fano} captures the essence of the quantum explanation. The important point is that one must consider quantum mechanical amplitudes for the detection of two photons. Since the two photons can come from different sources, an interference term alters the probability of detecting them depending on the relative path length differences between the sources and the detector. Later, Glauber carefully analysed multiple photon detection in the context of quantum electrodynamics and developed a powerful, coherent formulation of such problems that remains one of the cornerstones of modern quantum optics \cite{glauber}.

\begin{figure}[t] 
   \centering
   \includegraphics[width=7.0cm]{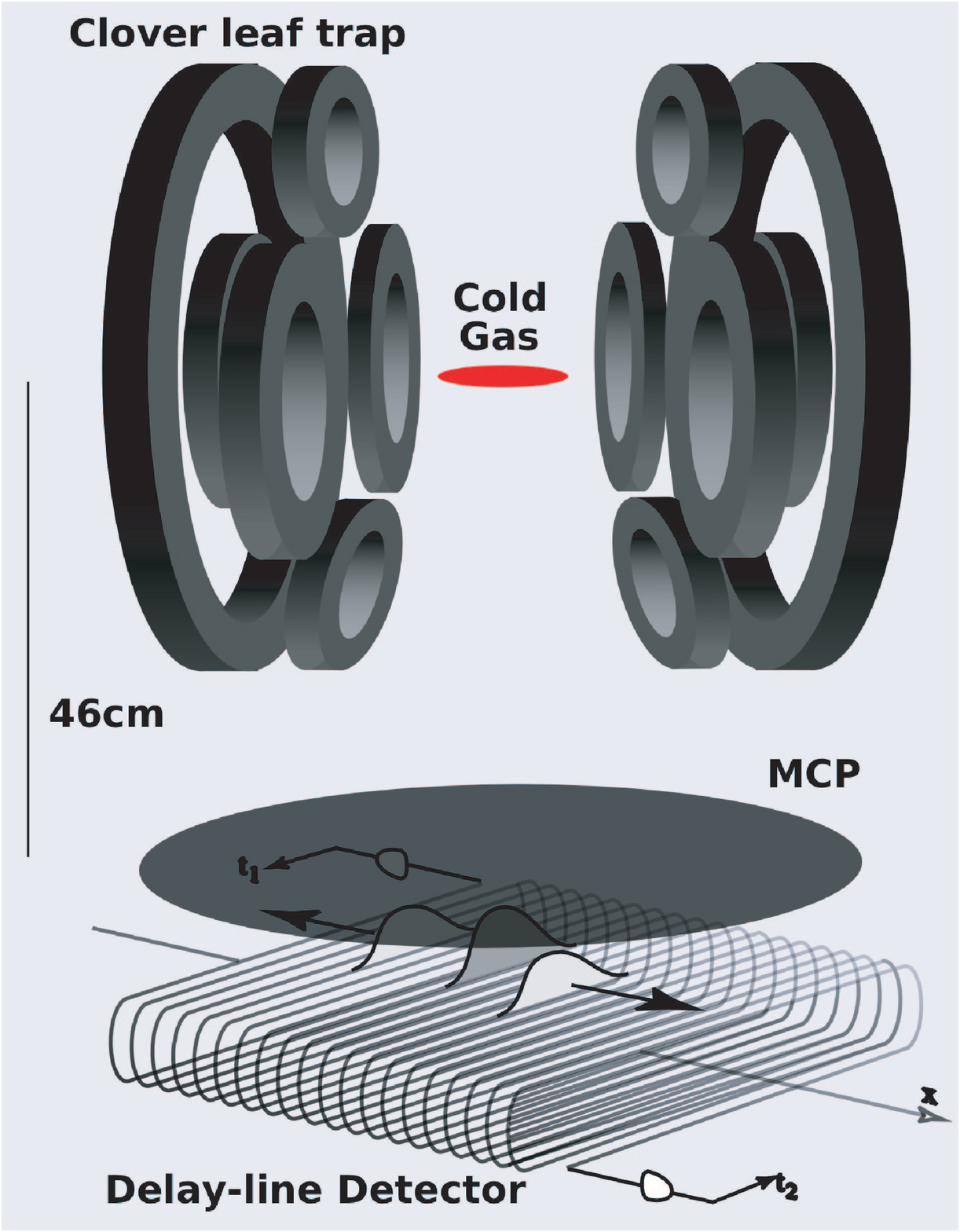} 
   \includegraphics[width=7.0cm]{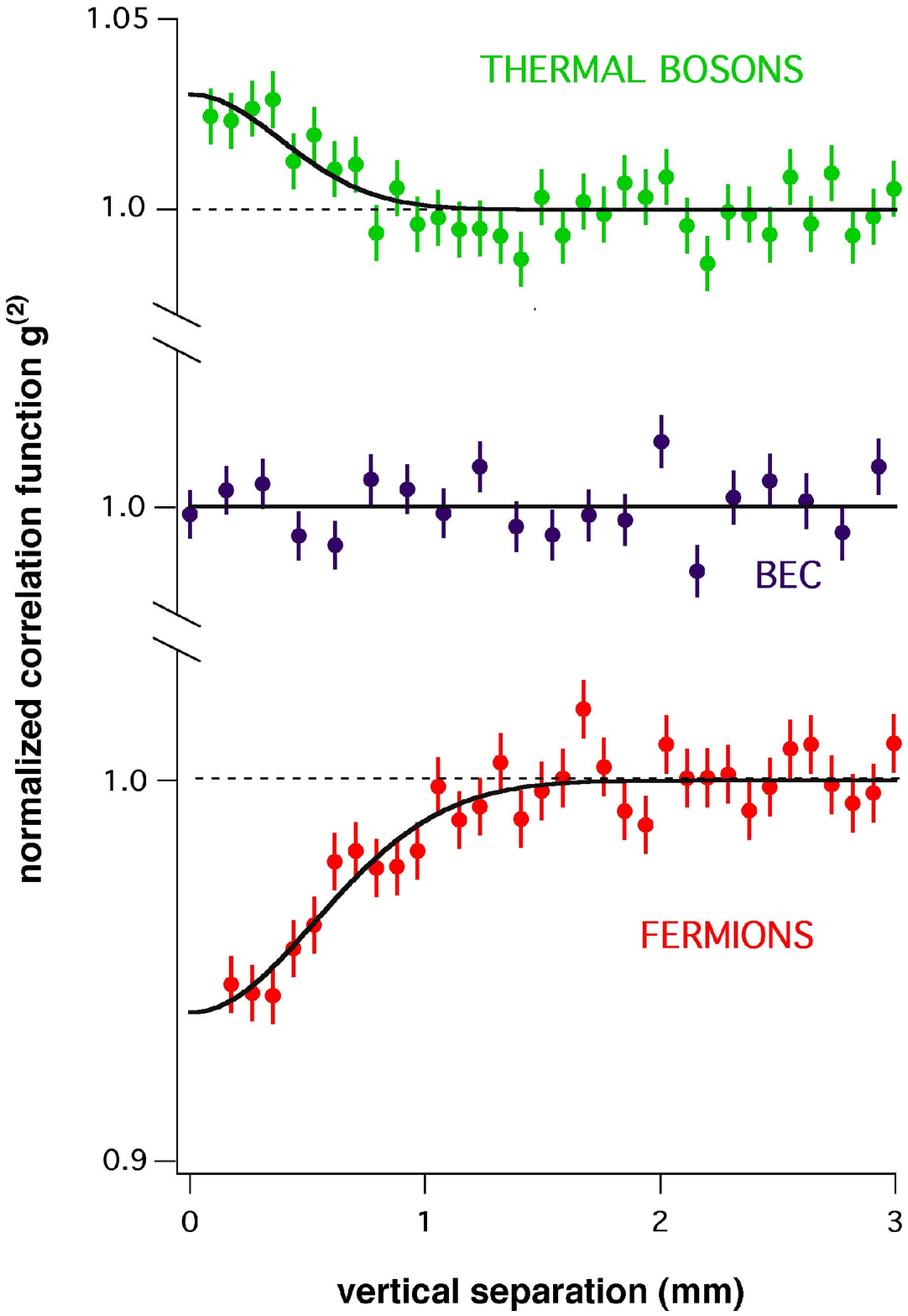} 
   \caption{{\it Left:} Schematic diagram of apparatus. A cloud of metastable helium atoms is formed at the center of a "cloverleaf" magnetic trap. The temperature is of order 1~$\mu$K. When the trap is switched off, the atoms fall onto a 8.5 cm diameter microchannel plate. A delay line anode and appropriate timing electronics permit the localization of individual atoms with 500~$\mu$m accuracy and 1 ns time resolution at the detector.
The pair correlation function is found from a histogram of pair separations.
{\it Right:} Normalized pair correlation functions for a thermal gas of bosons, a Bose-Einstein condensate and a thermal gas of fermions.
The correlation function has been averaged in the horizontal plane.
The vertical scale is the same for each data set.}
  \label{fig:apparatus}
\end{figure}

\section{Hanbury Brown Twiss with atoms}
	Since the appearance of the first atom interferometers in the 1990's the field of atom optics has made tremendous progress, often inspired by traditional optics \cite{cronin}. Recently it has become possible to realize fluctuation experiments analogous to the HBT experiment.  The first ones, carried out with bosons, give results highly analogous to those with photons. The results are of course unsurprising since photons are also bosons. Still, they nicely demonstrate this puzzling "mutual influence between particles", an influence which we have now learned to interpret as two particle quantum interference. In atom optics, one can also use fermions. The quadratic term in 
Eq.~\ref{eq:variance} appears with a minus sign, a manifestation of the exclusion principle, and it gives rise to an effect which has no optical or classical wave analog. 

	With these ideas in mind, our group at the Institut d'Optique has developed a detector capable of measuring spatio-temporal correlations between atoms by taking advantage of the properties of metastable helium. The metastable state $2^3S_1$ is 20 eV above the ground state and although its lifetime is 9000 s in vacuum, it rapidly deexcites when in contact with a metal surface (such as that of a microchannel plate) liberating 20 eV in the form of a free electron. The microchannel plate can thus provide an electrical pulse for a single atom. A position sensitive anode makes the plate the equivalent of an array of more than $10^4$ separate detectors, capable of  recording the arrival times and positions of a large number of atoms. The experimental set up is shown in figure~\ref{fig:apparatus}. A cloud of cold atoms falls, accelerated by gravity, onto the detector. With the three dimensional arrival information of each atom, one can construct the correlation function by histogramming the number of detected atom pairs as a function of their separation. In the vertical direction, the detector records arrival times, but since the atoms all travel at very nearly the same velocity, arrival times are easily converted to vertical positions. 
		
	In 2005 we used helium-4 (a boson) to observe the atomic HBT effect. In addition, by cooling the sample to below the Bose-Einstein condensation threshold, we were able to observe the absence of correlations in a BEC, illustrating the profound analogy between this state of matter and the light from a laser (figure~\ref{fig:apparatus} middle trace) \cite{schellekens}. More recently, in collaboration with a group in Amsterdam, we have also made the same measurement for helium-3, a fermion.  
In the fermion experiment we were able to study helium-3 and helium-4 clouds under nearly identical conditions in the same apparatus \cite{jeltes}. The clouds were of sufficiently low density that the comparison of the two isotopes in figure~\ref{fig:apparatus} (top and bottom traces) shows a purely quantum statistical effect. For thermal bosons the probability to detect two particles is increased while for fermions it is decreased. 

The spatial scale of the correlation can be understood with a calculation similar to that in the case of light \cite{gomes}; after a time of flight $t$, the correlation length is $h t / 2 \pi ms$, where $m$ is the atomic mass $h$ is Planck's constant and $s$ is the source size. The correlation length for light, $\lambda L / 2\pi s$ can be recovered by identifying $h / m v$ with the de Broglie wavelength of an atom moving at speed $v = L / t$. If the detector has arbitrarily good resolution, one expects a value 0 for fermions and 2 for bosons at zero separation. In the experiment, the amplitude of the signal is limited by the detector resolution resulting in a factor $g$ of order 15. In addition to our experiments at the Institut d'Optique, many related experiments have been performed in recent years \cite{yasuda,folling,greiner,ottl,esteve,rom}.

\section{Prospects}
In a sense, the correlations between atoms that we have observed are a consequence of elementary quantum theory, and one might ask why we and other researchers have gone to such efforts to observe them. One answer is that the experimental demonstration of non-trivial quantum phenomena, even elementary ones, often stimulates fruitful new ideas. But we also know that the demonstration of these correlations is only the beginning. The interaction between atoms, which we entirely neglect here, the possible formation of molecular dimers or the use of more complex configurations - such as putting the atoms in optical lattices \cite{folling,rom} - should lead to a rich variety of phenomena. We know that in lattices, confined in one or two dimensions or in rotating traps, ultra cold atoms constitute important testing grounds for models from condensed matter physics. Some even hope to shed light on phenomena which are still poorly understood such as high temperature superconductivity. 

	Hanbury Brown Twiss experiments have also been performed on other types of particles. In nuclear physics, correlations between pions give information about collision volumes in heavy ion collisions \cite{lisa}, thus realizing at the scale of femtometers a measurement analogous to HBT's measurements of stellar diameters. Conduction electrons in a solid form an essentially ideal fermi gas and HBT type anti-correlations have been observed \cite{electrons}. As in the case of atoms, these experiments are opening interesting paths towards the study of electron correlations in more exotic situations such as the fractional quantum Hall effect. 

\vskip 6pt
\noindent
{\bf Acknowledgements}
This work was funded by the CNRS, the ANR, the European Union and the Institut Francilien de Recherche sur les Atomes Froids.

\end{document}